# Spin Injection in Spin FETs Using a Step-Doping Profile

Min Shen, *Member IEEE*, Semion Saikin, Ming-Cheng Cheng, *Senior Member IEEE*

**Abstract— We investigate effect of a step-doping profile on the spin injection from a ferromagnetic metal contact into a semiconductor quantum well (QW) in spin FETs using a Monte Carlo model. The considered scheme uses a heavily doped layer at the metal/semiconductor interface to vary the Schottky barrier shape and enhance the tunneling current. It is found that spin flux (spin current density) is enhanced proportionally to the total current, and the variation of current spin polarization does not exceed 20%.**

## I. INTRODUCTION

Utilization of the electron spin as an information carrier in conventional semiconductor electronic devices results in a promising ideas for semiconductor spintronics [1-4]. Different types of spin-FETs [5-7] and bipolar spin-transistors [8-10] have been proposed. However, study of these devices is still at the early stage of development. One of most challenging problems of semiconductor spintronics is to produce spin-polarized currents in non-magnetic semiconductor structures. The conventional model of the spin injection from a ferromagnetic contact [11] utilized in the metal spintronics [1] is complicated by strong conductance mismatch between metal and semiconductor [12]. Injection through a tunneling barrier at the ferromagnetic metal/semiconductor interface has been suggested to resolve this problem [13]. Promising results of spin injection through different types of barriers have been reported recently [14-16]. In this paper, we study effect of a step-doping layer at the interface on spin injection through a Schottky barrier into a semiconductor QW.

Design of spintronic devices requires an appropriate shape of the Schottky barrier to achieve high spin injection. This can be realized by careful selection of material properties and variation of the doping profile to effectively control the spin-polarized current. For spin injection through a Schottky barrier, it was reported that the depletion region is detrimental due to strong and fast space-varying electric filed [17]. However, this effect can be minimized by one of the schemes using the barrier engineering [18], which is to use a high doping layer at the metal/semiconductor interface [16]. Efficient spin injection through a tailored Schottky barrier into a bulk semiconductor has been reported in [15,16].

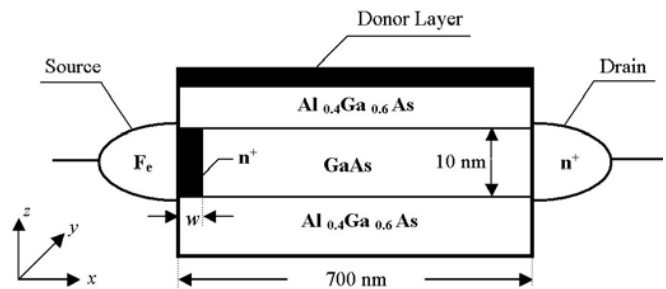

Fig. 1 Spin FET structure. The $n^+$ layer at the Fe/GaAs interface is used to vary the shape of the Schottky barrier. $w$ is the width of the high doping layer.

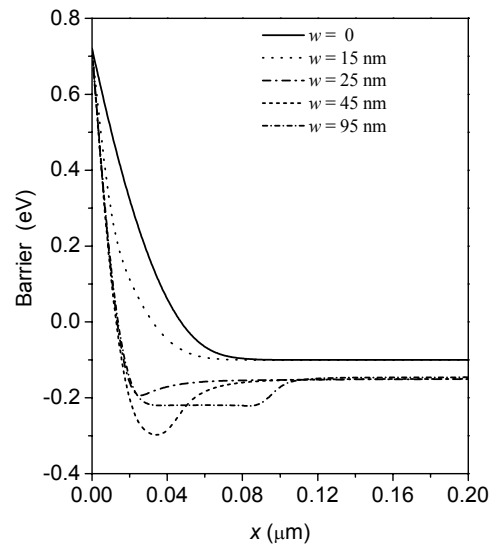

Fig. 2 Schottky barrier profiles for different widths of the heavily doped layer

In order to reveal and understand the effect of the step doping on spin injection, we apply the previously developed Monte Carlo scheme [19], which was used to study the spin

Manuscript received May 17, 2004. This research was supported by the National Security Agency and Advanced Research and Development Activity under Army Research Office contract DAAD-19-02-1-0035, and by the National Science Foundation grant DMR-0121146.

M. Shen and M.C. Cheng are with the Center for Quantum Device Technology and Department of Electrical & Computer Engineering, Clarkson University, Potsdam, NY 13699 USA (e-mail: shenm@clarkson.edu; mcheng@clarkson.edu).

S. Saikin is with the Center for Quantum Device Technology, and Department of Electrical & Computer Engineering, Clarkson University, Potsdam, NY 13699 USA. He is also with Physics Department, Kazan State University, Kazan 420008 Russia (e-mail: saikin@clarkson.edu).

This paper is based on the work presented at the 2004 IEEE NTC Quantum Device Technology Workshop.

injection through a Schottky barrier with a fixed doping profile. In this paper, we discuss the effect of barrier shape variation introduced by an additional heavily doped layer with $N_d = 2.5 \times 10^{24}$ cm$^{-3}$ at the metal/semiconductor interface, as shown in the spin-FET structure given in Fig. 1, similar to the spin FET proposed by Datta and Das [5]. Ferromagnetic metal, Fe, is used as the source contact, and the device channel is a QW of Al$_{0.4}$Ga$_{0.6}$As/GaAs/Al$_{0.4}$Ga$_{0.6}$As heterostructure. Heavily doped bulk GaAs is used in the drain with $N_d = 2.5 \times 10^{24}$ cm$^{-2}$. This study focuses on the effect of the high doping layer at the metal/semiconductor interface on the spin injection. Collection of electrons in the drain is assumed spin-independent.

## II. MODEL

The Monte Carlo model, described in [19], takes into account thermionic emission and tunneling mechanism from metal to semiconductor and from semiconductor to metal. The barrier height is assumed to be 0.72 eV [20] and bias-independent. The quantum well depth is approximately 0.35 eV [21], and the QW width is 10 nm.

Both Rashba [22] and Dresselhaus [23] effects are included in the spin orbit interaction, which are described by

$$H_R = \eta(\sigma_x k_y - \sigma_y k_x) \tag{1}$$

and

$$H_D = \beta[(\langle k_z^2 \rangle - k_x^2)\sigma_x k_y - (\langle k_z^2 \rangle - k_y^2)\sigma_x k_x], \tag{2}$$

respectively. $\eta$ and $\beta$ are Rashba and Dresselhaus spin-orbit coupling coefficients, respectively. For GaAs, we use the calculated value, $\beta = 28$ eV·Å$^3$ [24], while $\eta = 0.005$ eV·Å is comparable with the measured value [25]. In Eqs. (1) and (2), the coordinate system coincides with the principal crystal axes. The single-electron density matrix is used to describe spin evolution. The evolution of spin density matrix $\rho$ is performed as

$$\rho(t + \Delta t) = e^{-iH_{SO}\Delta t/\hbar} \rho(t) e^{iH_{SO}\Delta t/\hbar}, \tag{3}$$

where

$$H_{SO} = H_R + H_D. \tag{4}$$

To describe the spin injection, we use spin current density (or spin flux) defined as

$$J_{\sigma_\alpha}^\beta = \sum_i v_\beta^i Tr(\sigma_\alpha \rho_i), \tag{5}$$

where $v_\beta^i = \hbar k_\beta^i / m^*$ is the $\beta$-component velocity of the $i$-th electron and $\sigma_\alpha$ is the Pauli matrix. Effect of the spin-orbit splitting on the wave vector **k** is assumed negligible. In the spin-independent case, the E-**k** relation then reduces to the conventional one. If only the linear spin orbit interaction in momentum is included in Eqs. (1) and (2), it is similar to the E-**v** relation [26]. If the cubic terms in the Dresselhaus interaction are taken into account in Eq. (2), this approximation ignores the **k** broadening of the single-electron wave packet. For spin polarized currents (but not for pure spin currents [27,28]), the normalized current spin polarization

$$P_{\sigma_\alpha}^{J_\beta} = J_{\sigma_\alpha}^\beta / J_\beta \tag{6}$$

can be introduced, where $J_\beta$ is the $\beta$-component of the total current density. In general, this characteristic of current spin polarization differs from the particle spin polarization used in [29, 30]. We found it to be useful for studying spin dynamics. In the following text, we discuss the spin-polarized current along external electric field applied in the $x$ direction. Therefore, the notations for spin current density, total current density and current spin polarization are simplified as, $J_{\sigma_\alpha} \equiv J_{\sigma_\alpha}^x$, $J \equiv J_x$ and $P_\alpha \equiv P_{\sigma_\alpha}^{J_x}$, respectively. For absolute values, we use $J_s = \sqrt{\sum_\alpha (J_{s\alpha})^2}$ and $P = \sqrt{\sum_\alpha (P_\alpha)^2}$. Because a total spin in the system with spin-orbit interaction is not conserved, spin current and spin current polarization are coordinate dependent.

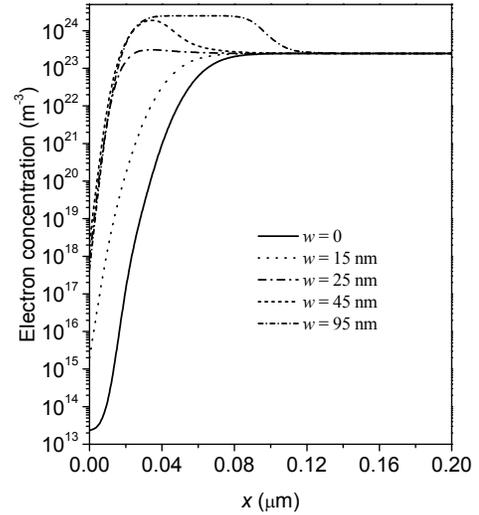

Fig. 3  Electron concentrations for different widths of the heavily doped layer

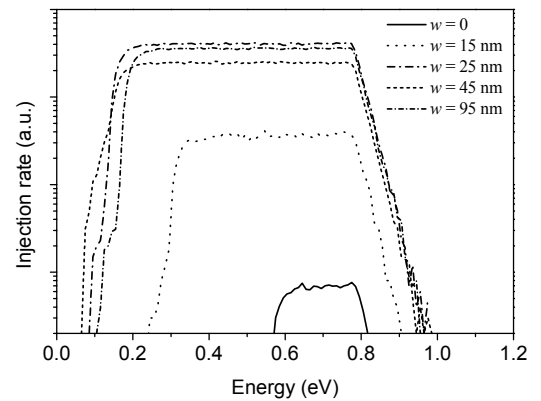

Fig. 4  The distribution of the injection rate vs. energy for different widths of the heavily doped layer.

### III. SIMULATION RESULTS AND DISCUSSION

Simulations are performed at room temperature in the structure given in Fig. 1 with $w = 0$, 15, 25, 45 and 95 nm at the source-drain voltage $V_{DS} = 0.1$ V. The barrier profiles and electron concentrations for five different widths of the high doping layer, determined from the self-consistent solution of Poisson equation and electron motion, are shown in Figs. 2 and 3. This naturally incorporates effects of inhomogeneous doping on spin dynamics [31]. Inclusion of the heavily doped layer at the contact interface narrows the barrier width. The decrease in the barrier width with $w$ however becomes saturated at $w \approx 30$ nm. It should be mentioned that Fig. 2 shows the conduction band profiles only in the region of $0 < x < 200$ nm while the channel extends to $x = 700$ nm, as displayed in Fig. 1. Though in the channel the conduction band profiles vary with $w$, as shown in Fig. 2, these band energies derived from different values of $w$ eventually converge to the same value at the $n^+$ drain.

Fig. 4 shows the influence of the change in the barrier profile on the energy distribution of the injected electrons. The distribution at the higher energy edge is controlled by the barrier height (i.e., dominated by the thermionic emission) while the distribution at the lower energy edge is influenced by the tunneling efficiency. The area under the distribution reflects the injection strength. According to Fig. 4, introduction of the interface step-doping layer increases the injection strength. In addition, the threshold energy level for the evident injection rapidly decreases with the layer width $w$, as $w$ increase from 0. At the considered bias, $V_{DS} = 0.1$ V, the threshold energy reach however a minimum level near 0.1 eV for $w > 25$ nm. Fig. 2 shows that the conduction band energy decreases rapidly with $x$ near the Schottky contact, especially for $w \neq 0$. The grid size ($\Delta x = 0.01$ μm) used in the simulation does not provide enough spatial resolution near the contact. It is believed that the insufficient spatial resolution leads to a tunneling probability which approximately increases exponentially with energy and compensates the exponentially decreasing Maxwellian distribution function of electrons in the metal. As a result, a nearly constant energy distribution of the injection rate is observed within the energy range where the spatial resolution is insufficient.

Figs. 5(a) and 5(b) show three components of spin flux for the spin injection in the structure without the interface step-doping layer ($w = 0$). Injected electrons are 100% spin polarized in the $x$ and $y$ orientations. Both, linear and nonlinear spin orbit interactions, given in Eqs. (1) and (2) are included. This case is used as the reference to study effects of the barrier profile on spin dynamics.

The current spin polarizations defined in Eq. (6) are illustrated in Figs. 6(a) and 6(b), corresponding to the cases presented in Figs. 5(a) and 5(b), respectively. Results accounting for only linear spin orbit interaction are also displayed in Figs. 6(a) and 6(b) as solid lines. In this case, the small difference in the spin polarization profiles along the channel for different polarizations of injected electrons results from the anisotropy of spin relaxation rates in semiconductor heterostructures [32, 29]. Although in most studies of spin FETs only the linear spin-orbit terms are included, Fig. 6 shows that, in the considered structure, current spin polarization is strongly influenced by the nonlinear Dresselhaus term. The current spin component in the channel direction decays on a length scale of 0.1 μm when the nonlinear term is included. Though, the $y$ component of the current spin polarization relaxes appreciably more slowly. We attribute this strong anisotropy of spin dynamics to the velocity distribution of the injected electrons rather than the interplay of the Rashba and Dresselhaus coefficients.

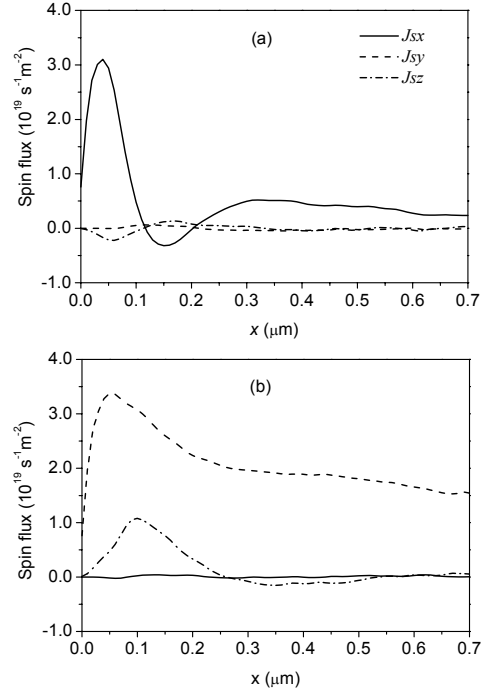

Fig. 5 Three components of spin flux with spin-polarized injection (a) in the $x$ orientation and (b) in the $y$ orientation

Our following discussion about the width of the step-doping effect will be focused on the injection spin polarization in the $y$ direction in which spin polarization is conserved in a much longer length scale. The interface doping tailors the barrier profile that induces the following effects:

1) change in the tunneling probability and thus tunneling current,
2) modification of the energy distribution of the injected particles, which influences spin dynamics in the channel,
3) change in the initial distribution of spin polarization in the case of non-100% spin-polarized injection.

Attempts are made to analyze these three effects separately.

Fig. 7 shows the ratio of the injected current density in the channel in the case with a step-doping layer to that with no step doping (the reference case shown in Fig. 5) as a function of the layer width. The absolute value of the maximum variation for the current spin polarization induced by the various widths of the step-doping layer with respect to the reference case, $\Delta P(x)$, is shown in the inset, where

$$\Delta P(x) = \max_w \left( \frac{P(w,x) - P^{\text{ref}}(x)}{P^{\text{ref}}(x)} \right). \qquad (7)$$

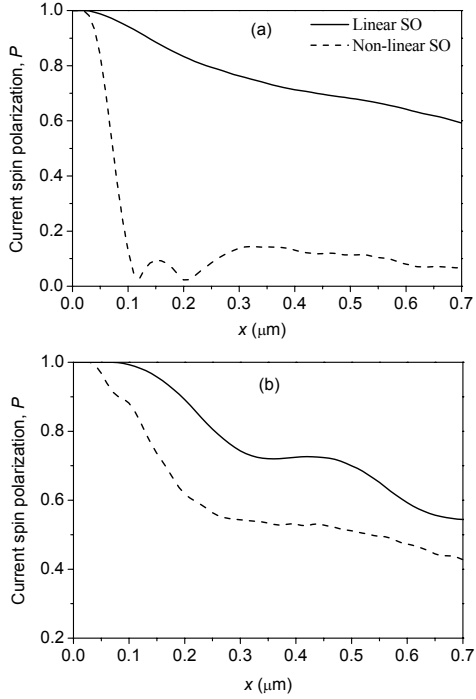

Fig. 6 Current spin polarization with spin-polarized injection (a) in the *x* orientation and (b) in the *y* orientation

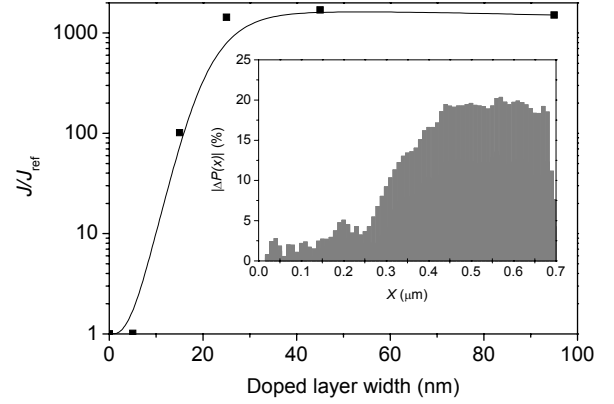

Fig. 7 The ratio of total injected current density in the channel direction for the case with the high doping layer to the reference case (no doping layer) versus the layer width. The absolute value of the maximum variation of the current spin polarization (caused by the doped layer of various widths) with respect to the reference case is shown in the inset.

$\Delta P(x)$ changes from 2.5% up to 20% of the total polarization along the channel. Effect of different widths of the step-doping layer on the current polarization actually does not exceed 20%. However, the spin current strongly depends on the doping layer width due to the enhancement of total current density. The current density (therefore the spin current density) is exponentially dependent on $w$ for $w < 25$ nm, as shown in Fig. 7. The ratio is saturated for $w > 30$ nm, which is consistent with the saturation of the barrier thinning shown in Fig. 3.

The above discussion is based on a 100% polarization for the injected electrons. Realistically, electrons are injected from the ferromagnetic contact with a certain polarization $P(E)$ that is dependent on electron energy, $E$, material parameters, and interface quality, etc. It can be derived based on first principle calculations [33]. For simplicity, to check the effect of the initial polarization on the spin dynamics in the device, we approximate $P(E)$ by the ratio of densities of states between the majority and minority spins in the metal contact [19]. Fig. 8 presents the deviation of the absolute percentage variation for the current spin polarization due to the non-100% spin-polarized injection with respect to the reference case (100% spin-polarization). In both cases, $w = 15$ nm. The injection efficiency with respect to the reference case shifted to $\Delta P(x = 0) = 42\%$, and the variation along device channel is about 8%. This indicates that effects of the initial polarization, $P(E)$, and the barrier profile on spin polarized current are nearly separable.

There are other parameters, such as the step-doping density and the barrier height, which can be adjusted to improve the results. This will be studied in the near future.

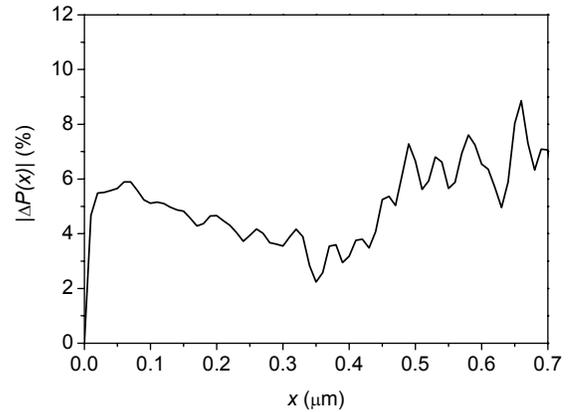

Fig. 8 The shifted variation of the current spin polarization in the channel direction.

## IV. CONCLUSION

We study the effect of different widths of the step-doping layer at the metal-semiconductor interface on the spin injection in spin FETs. The Monte Carlo simulation results indicate that the higher-order spin-orbit term plays an important role in spin injection transport. It is found that spin flux (spin current density) is enhanced exponentially by an increase in the width of the step doping at the contact-semiconductor interface for $w < 25$ nm, and becomes saturated for $w > 30$ nm. The influence of the width variation of the high step-doping layer on the current spin polarization actually does not exceed 20%.

## ACKNOWLEDGEMENT

The authors thank Prof. V. Privman for valuable discussions.

## REFERENCES

[1] S. A. Wolf, D. D. Awschalom, R. A. Buhrman, J. M. Daughton, S. von Molnar, M. L. Roukes, A. Y. Chtchelkanova, and D. M. Treger, "Spintronics: A spin-based electronics. vision for the future," *Science*, vol. 294, pp. 1488–1495, 2001.

[2] I. Zutic, J. Fabian and S. Das Sarma, "Spintronics: Fundamentals and applications", Rev. Modern Phys., vol. 76, pp 323-410, 2004.

[3] D. D. Awschalom, M. E. Flatte, and N. Samarth, "Spintronics," *Sci. Amer.*, vol. 286, no. 6, pp. 66–73, 2002.

[4] H. Akinaga, H. Ohno, "Semiconductor spintronics," *IEEE Trans. Nanotechnol.*, vol. 1, no. 1, pp. 19–31, 2002.

[5] S. Datta and B. Das, "Electronic analog of the electro-optic modulator," *Appl. Phys. Lett.*, vol. 56, pp. 665–667, 1990.

[6] J. Schliemann, J. C. Egues, and D. Loss, "Non-ballistic spin-field-effect transistor," *Phys. Rev. Lett.*, vol. 90, art. no. 146801, 2003.

[7] J. C. Egues, G. Burkard, and D. Loss, "Datta–Das transistor with enhanced spin control," *Appl. Phys. Lett.*, vol. 82, pp. 2658–2660, 2003.

[8] J. Fabian, I. Zutic, S. D. Sarma, "Theory of spin-polarized bipolar transport in magnetic p-n junctions," *Phys. Rev. B*, vol. 66, art. no. 165301, 2002.

[9] M. E. Flatte, Z. G. Yu, E. Johnson-Halperin, D. D. Awschalom, "Theory of semiconductor magnetic bipolar transistors," *Appl. Phys. Lett.*, vol. 82, pp. 4740–4742, 2003.

[10] J. Fabian, I. Zutic, S. D. Sarma, "Magnetic bipolar transistor," *Appl. Phys. Lett.*, vol 84, pp. 85–87, 2004.

[11] P. R. Hammar, B. R. Bennett, M. J. Yang, M. Johnson, "Observation of spin injection at a ferromagnet-semiconductor interface," *Phys. Rev. Lett.*, vol. 83, pp. 203–206, 1999.

[12] G. Schmidt, D. Ferrand, L. W. Molenkamp, A. T. Filip, B. J. van Wees, "Fundamental obstacle for electrical spin injection from a ferromagnetic metal into a diffusive semiconductor," *Phys. Rev. B*, vol. 62, pp. R4790–R4793, 2000.

[13] E. I. Rashba, "Theory of electrical spin injection: Tunnel contacts as a solution of the conductivity mismatch problem," *Phys. Rev. B*, vol. 62, pp. R16267–R16270, 2000.

[14] K. H. Ploog, "Spin injection ferromagnet-semiconductor heterostructures at room temperature", *J. Appl. Phys.*, vol. 91, pp7256–7260, 2002.

[15] V. F. Motsnyi, J. De Boeck, J. Das, W. Van Roy, G. Borghs, E. Goovaerts, V. I. Safarov, "Electrical spin injection in a ferromagnet/ tunnel barrier/semiconductor heterostructure," *Appl. Phys. Lett.*, vol. 81 pp. 265–267, 2002.

[16] A. T. Hanbicki, O. M. J. van 't Erve, R. Magno, G. Kioseoglou, C. H. Li, B. T. Jonker, G. Itskos, R. Mallory, M. Yasar, A. Petrou, "Analysis of the transport process providing spin injection through an Fe/AlGaAs Schottky barrier," *App. Phys. Lett.*, vol. 82, pp 4092–4094, 2003.

[17] J.D. Albrecht, D.L. Smith, "Electron spin injection at a Schottky contact," *Phys. Rev. B*, vol. 66, art. no. 113303, 2002.

[18] S. Sassen, B. Witzigmann, C. Wolk, H. Brugger, "Barrier height engineering on GaAs THz Schottky diodes by means of high-low doping, InGaAs- and InGaP-Layers," *IEEE Trans. Devices*, vol. 47, pp 24–32, 2000.

[19] M. Shen, S. Saikin, M.-C. Cheng, "Monte Carlo modeling of spin injection through the Schottky barrier and spin transport in a semiconductor Quantum Well," http://arxiv.org/abs/cond-mat/0405270

[20] J. R. Waldrop, "Schottky barrier height of ideal metal contacts to GaAs," *Appl. Phys. Lett.*, vol. 44, pp. 1002–1004, 1984.

[21] Website: www.ioffe.rssi.ru/SVA/NSM/ Semicond/AlGaAs/bandstr.html

[22] Yu. Bychkov, E. I. Rashba, "Oscillatory effects and the magnetic susceptibility of carriers in inversion layers," *J. Phys. C*, vol. 17, pp. 6039–6045, 1984.

[23] G. Dresselhaus, "Spin-orbit coupling effects in Zinc Blende structures," *Phys. Rev.*, vol. 100, pp. 580–586, 1955.

[24] M. Cardona, N. E. Christensen, G. Fasol, "Relativistic band structure and spin-orbit splitting of zinc-blende-type semiconductors," *Phys. Rev. B*, vol. 38, pp. 1806–1827, 1988.

[25] J. B. Miller, D. M. Zumbühl, C. M. Marcus, Y. B. Lyanda-Geller, D. Goldhaber-Gordon, K. Campman, A. C. Gossard, "Gate-controlled spin-orbit quantum Interference effects in lateral transport," Phys. Rev. Lett. vol. 90, art. no. 076807, 2003.

[26] Bandy. S. Pramanik and S. Bandyopadhyay, M. Cahay, "Decay of spin-polarized hot carrier current in a quasi-one-dimensional spin-valve structure," Applied Phys. Lett., vol. 84, pp. 266–268, 2004.

[27] Shuichi Murakami, Naoto Nagaosa, Shou-Cheng Zhang, "Dissipationless quantum spin current at room temperature," *Science*, vol. 301, pp. 1348–1351, 2003.

[28] J. Sinova, D. Culcer, Q. Niu, N. A. Sinitsyn, T. Jungwirth, A. H. MacDonald, "Universal Intrinsic Spin Hall Effect," *Phys. Rev. Lett.*, vol. 92, art. no. 126603, 2004.

[29] S. Pramanik, S. Bandyopadhyay, M. Cahay, "Spin dephasing in quantum wires," *Phys. Rev. B*, vol. 68, art. no. 075313, 2003.

[30] S. Saikin, M. Shen, M.-C. Cheng, V. Privman, "Semiclassical Monte Carlo model for in-plane transport of spin-polarized electrons in III-V heterostructures," *J. Appl. Phys.*, vol. 94, pp. 1769–1775, 2003.

[31] Y. V. Pershin, V. Privman, "Focusing of spin polarization in semiconductors by inhomogeneous doping," *Phys. Rev. Lett.*, vol. 90, art. no. 256602, 2003.

[32] Kach. M. I. Dyakonov and V. Y. Kachorovskii, "Spin relaxation of two-dimensional electrons in noncentrosymmertic semiconductors," *Sov. Phys.Semicond.*, vol. 20, pp. 110–112, 1986.

[33] X.-G. Zhang, W.H. Butler, "Band structure, evanescent states, and transport in spin tunnel junctions," *J. Phys.: Cond. Matt.*, vol. 15, pp. R1603–R1639, 2003.